\documentclass[aps, twocolumn, 8pt]{revtex4}
\usepackage{latexsym}
\usepackage{graphicx}
\usepackage{amssymb}

\def\RR{\mathbb R}

\begin{document}

\title{Sequential Attack with Intensity Modulation on the Differential-Phase-Shift Quantum Key Distribution Protocol}
\author{Toyohiro Tsurumaru}       %The name of the first author
\affiliation{Mitsubishi Electric Corporation,\\
Information Technology R\&D Center\\
5-1-1 Ofuna, Kamakura-shi, Kanagawa,
247-8501, Japan}

\begin{abstract}
In this paper,
we discuss the security of the differential-phase-shift quantum key distribution (DPSQKD) protocol
by introducing an improved version of the so-called sequential attack,
which was originally discussed by Waks et al.\cite{Waks06}.
Our attack differs from the original form of the sequential attack in that the attacker Eve modulates
not only the phases but also the amplitude in the superposition of the single-photon states which she sends to the receiver.
Concentrating especially on the ``discretized gaussian" intensity modulation,
we show that our attack is more effective than the individual attack, which had been the best attack up to present.
As a result of this,
the recent experiment with communication distance of 100km reported by Diamanti et al.\cite{Diamanti06}
turns out to be insecure.
Moreover it can be shown that in a practical experimental setup which is commonly used today,
the communication distance achievable by the DPSQKD protocol is less than 95km.
\end{abstract}

\maketitle

\section{Introduction}
The differential-phase-shift quantum key distribution (DPSQKD) protocol\cite{Inoue02}
is a promising protocol for quantum key distribution (QKD) featuring the tolerance against
photon number splitting attacks.
However, its security has only been investigated so far against limited types of attacks;
The most effective attack up to present had been a particular type of individual attacks
investigated by Waks et al. \cite{Waks06},
in which Eve acts on {\it photons} individually rather than on {\it signals}.
Several long-distance experiments claiming security along its line have been reported (see, e.g.,\cite{Diamanti06} and references therein).
Another important class of attacks is a variant of intercept-resend attacks called {\it sequential attacks}
which was also discussed originally by Waks et al.\cite{Waks06} and later improved by Curty et al.\cite{CZLN06}.
However, they still remain less effective than the individual type for most probable applications.

We will here present a further improvement on the sequential attack
in which the attacker Eve modulates not only the phases but also
the amplitude in the superposition of the single-photon states which she sends to the receiver.
In this paper,
with a slight abuse of terminology,
we call such differentiations in the amplitude an {\it intensity modulation} and investigate its consequence.
Concentrating especially on the ``discretized gaussian" intensity modulation,
we show that our attack is more effective than the individual attack, which had been the best attack.
As a result of this,
the recent experiment with communication distance of 100km reported by Diamanti et al.\cite{Diamanti06}
turns out to be insecure.
Moreover it can be shown that in a practical experimental setup which is commonly used today,
the communication distance achievable by the DPSQKD protocol is less than 95km.

In what follows, we shall consider a conservative definition of security, i.e.,
we assume that Eve can control some flaws in Alice's and Bob's devices
(e.g., the detection efficiency and the dark count probability of the detectors),
together with the losses in the channel, and she exploits them to obtain
maximal information about the shared key.

\section{DPSQKD protocol}
\begin{figure}[ht]
\begin{center}
\includegraphics[trim=0cm 7cm 1.5cm 3.5cm, clip, scale=0.35]{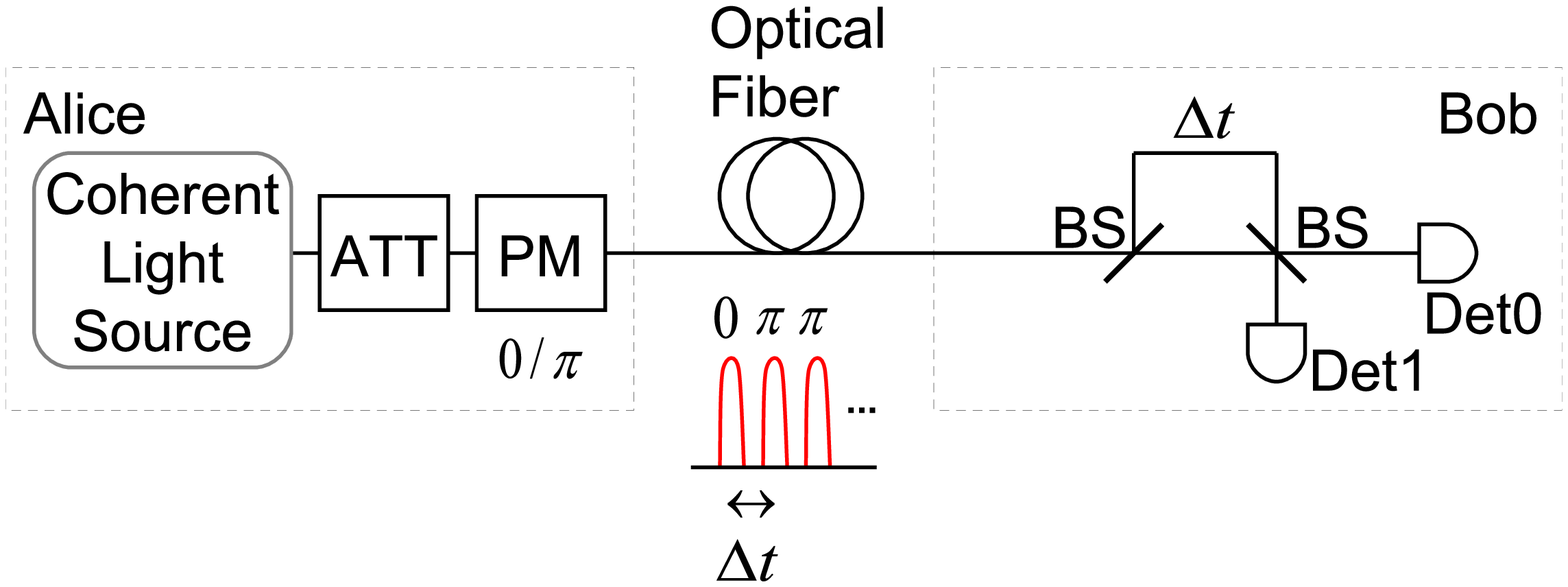}
\end{center}
\caption{Basic experimental setup of a DPSQKD system.
Here PM denotes a phase modulator, and BS a 50:50 beamsplitter.
$\Delta t$ represents the time difference between two consecutive pulses.
Photon detectors Det0 and Det1 output the phase differece $z_i=0$ and $1$ respectively.}
\label{fig:DPSQKD_scheme}
\end{figure}

The DPSQKD protocol proceeds as follows\cite{Inoue02,Waks06}.
\begin{enumerate}
\item
Alice generates a random bit string $x=(x_0,\dots,$
$x_N)$, $x_i\in\{0,1\}$
and sends to Bob the coherent light pulses with phase
$\phi_i=\pi x_i+\phi$ and intensity $\alpha$,
where $\phi,\alpha\in\mathbb{R}$, $\alpha>0$.
That is, Alice generates the following $|\Psi\rangle$ and sends to Bob:
\[
|\Psi\rangle=|\alpha e^{i\phi_0}\rangle\otimes\cdots\otimes|\alpha e^{i\phi_N}\rangle.
\]
Intensity $\alpha$ is related to the average photon number $\bar{n}$ per pulse
as $\bar{n}=|\alpha|^2$.

\item
Using a Mach-Zehnder interferometer (Fig.\ref{fig:DPSQKD_scheme}),
Bob measures the phase difference $z_i=x_i+x_{i-1}$ of each adjacent pair of pulses\footnote{Throughout
this paper, summations over $x_i$, $z_i$, $b_i$ are always in modulo 2.}.
The outcome of the measurement is recorded as the sifted key $b=(b_1,\dots,b_l)$,
 $b_a=x_{i_a}+x_{i_a-1}$, where $i_1$,$\dots$,$i_l$ are the times when photons are detected.
\item Communicating on an authenticated classical channel,
Alice and Bob perform key reconciliation and the generalized privacy amplification\cite{BBCM95,Waks06}
to calculate the secret key.
\end{enumerate}
We denote by $T$ the transmission of the communication channel including
the loss in fiber and Bob's interferometer, and the quantum efficiency of Bob's detector.
We also denote by $d$ the dark count of Bob's detector.
Then the signal and detection rate $p_{\rm click}$ at Bob's side is given by
\begin{equation}
p_{\rm click}=T\bar{n}+d.
\label{eq:detection_prob}
\end{equation}
Thus in the second step above, the average length $l$ of the sifted key is $l=Np_{\rm click}$.

%%%%%%%%%%%%%%%%%%%%%%%%%%%%%%%%%%%%%%%%%%%%%%%%%%%%%%%%%%%%%%%%%%%%
%%%%%%%%%%%%%%%%%%%%%%%%%%%%%%%%%%%%%%%%%%%%%%%%%%%%%%%%%%%%%%%%%%%%

\section{Sequential Attack with Intensity Modulation}
In this section,
we first define our attack
and then calculate several probabilities that are necessary for later analyses,
i.e., the error rate which Bob detects as a result of the attack and the success probability of Eve's attack.
The necessary conditions for Eve to carry out the attack are also derived in terms of experimental parameters,
such as the average photon number $\bar{n}$ and the transmission $T$.

\subsection{Description of Attack}
\begin{figure}[ht]
\begin{center}
\includegraphics[trim=1cm 7cm 0.5cm 0.5cm, clip, scale=0.35]{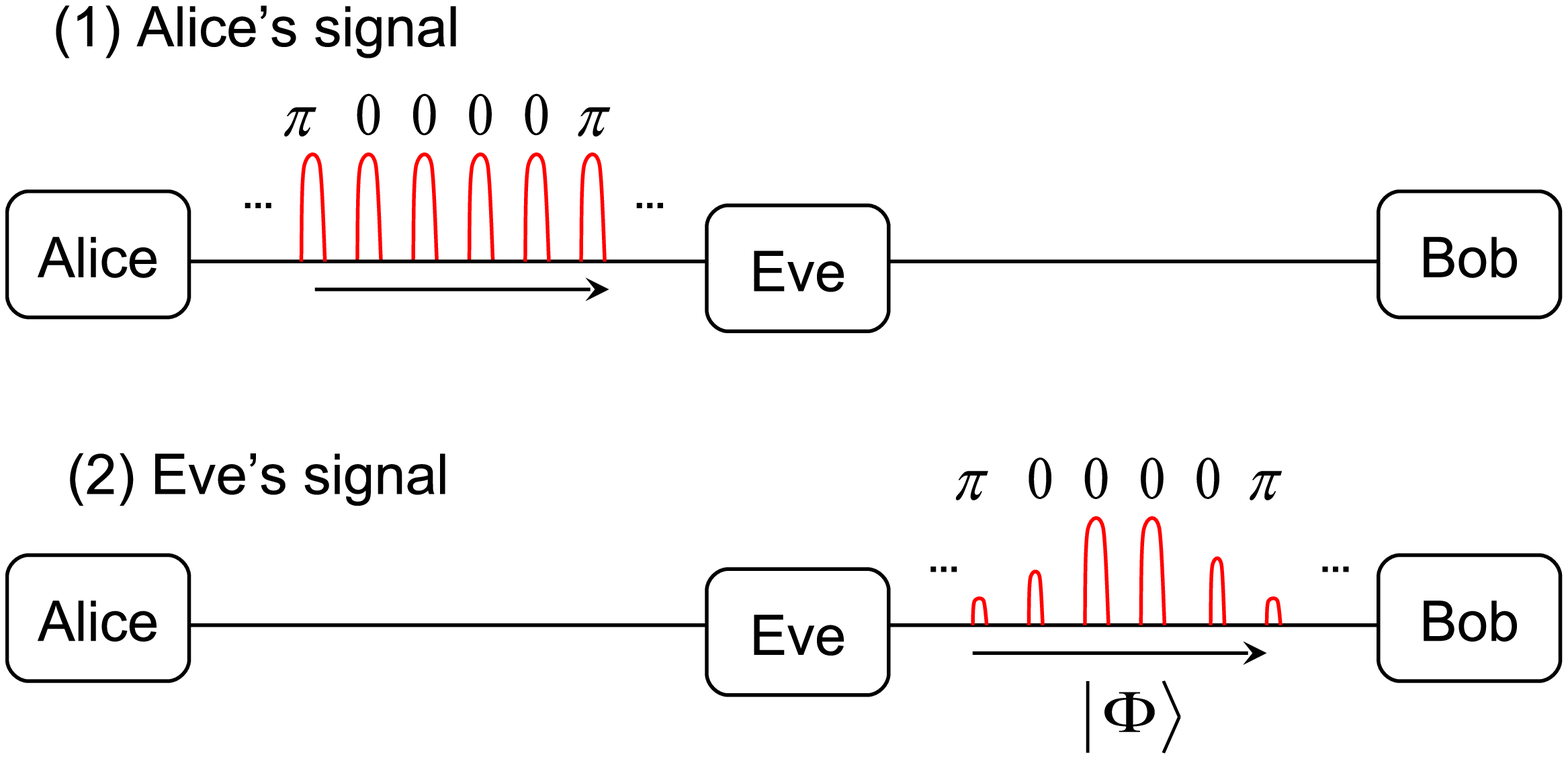}
\end{center}
\caption{Graphical image of our sequential attack.
(1)Alice emits coherent light pulses with an equal intensity,
and Eve detects them.
(2)When sequential detection events are found,
Eve sends out single-photon state $|\Phi\rangle$ of Eqn.(\ref{eq:defPhi}),
which is a superposition of single-photon states having the same phases as Alice's original signal
but with different amplitudes.}
\label{fig:attack}
\end{figure}

The basic scheme of our sequential attack is the same as the previous versions given in Ref.\cite{Waks06}.
The only difference is that the state $\Psi$ which Eve sends is
a superposition of single-photon states with different amplitudes (see Fig.\ref{fig:attack}).
The precise definition of our attack is as follows:
\begin{enumerate}
\item Using unambiguous state discrimination (USD, see Ref.\cite{CZLN06}) measurement,
Eve measures the phase $x_i$ of all but small portion of the pulses that Alice sent
and records the outcome.

Here the small portion means a pulse sequence of length $\epsilon N$ with $\epsilon=(1-r)p_{\rm click}/\bar{n}$.
The ratio $r\in \RR$, $0<r\le1$ is a parameter to be specified later.
\item For $rp_{\rm click}N$ sequential detection events of length $k$ or larger,
Eve does as follows:

If the detected phases were, say, $x_a,\dots,x_{a+k-1}$,
she generates state $|\Phi\rangle$ defined as
\begin{equation}
|\Phi\rangle:=\sum_{i=a-M}^{a+k+M}(-1)^{y_i}A_ia_i^\dagger|0\rangle,
\label{eq:defPhi}
\end{equation}
and sends it to Bob.
Here parameter $M\in \mathbb{N}$ is a sufficiently larege number,
and $y_i$ are set as $y_i=x_i$ for $a\le i\le a+k-1$, and other $y_i\in\{0,1\}$'s are randomly chosen values.
The creation operator $a_i^\dagger$ denotes a single photon pulse incident to Bob
for time interval $i$.
\item Eve sends $\epsilon N$ pulses which she kept intact in the first step, and sends them to Alice.
\end{enumerate}
In what follows, we assume that transmission $T$ and $p_{\rm click}/\bar{n}$ are small enough
(e.g., with the communication distance being sufficiently large)
so that $\epsilon$ can be neglected.

%%%%%%%%%%%%%%%%%%%%%%%%%%%%%%%%%%%%%%%%%%%%%%%%%%%%%%%%%%%%%%%%%%%%
%%%%%%%%%%%%%%%%%%%%%%%%%%%%%%%%%%%%%%%%%%%%%%%%%%%%%%%%%%%%%%%%%%%%

\subsection{Probability of Sequential Detection}
The USD measurement performed in the first step succeeds for each pulse with probability
\[p_{\rm USD}(\bar{n})=1-\exp(-2\bar{n}),\]
where $\bar{n}=|\alpha|^2$ is the average number of photon per pulse\cite{CZLN06}.
It can easily be shown that when Eve repeats it for a sufficiently long sequence of pulses,
a sequential detection with length $k$ or larger occurs with probability
\begin{equation}
p_{\rm seq}(k,\bar{n})=\left(1-p_{\rm USD}(\bar{n})\right)\left(p_{\rm USD}(\bar{n})\right)^k.
\label{eq:sequential_det}
\end{equation}
In other words, as a result of USD measurements on $N$ pulses,
Eve finds on average $Np_{\rm seq}(k,\bar{n})$ events of sequential detections of at least length $k$.
Hence $rp_{\rm click}\le p_{\rm seq}(k,\bar{n})$ is necessary in order for Eve to carry out her attack.

\subsection{Error Rate Due to $\Phi$}
Every state $|\Phi\rangle$ sent by Eve in the second step will always yield one detection event in Bob's detector in either of time $i$.
In other words, Bob's overall detection probability of $|\Phi\rangle$ is strictly one.
However, the detection can generally occur in a wrong detector causing the bit flip in the sifted key,
or it may click in time $i$ where Eve does not know the corresponding phase shift,
in which case she fails to steal the sifted key bit.

To evaluate these effects we will below obtain,
in terms of the amplitude $A_i$, the formulae for the error rate $E(k)$
caused by each $|\Phi\rangle$, and the probaility $D(k)$ that Eve ends up reading the sifted key bit using $|\Phi\rangle$.
In what follows, we suppose that $N$ is sufficiently large and $A_i$ converges to zero fast enough for $i\to\pm\infty$,
so that parameter $M$ introduced in the second step of our attack can be regarded as infinite .

Suppose that Eve has succeeded in determining $k$ subsequent phases $x_a,\dots,x_{a+k-1}$ in the first step
and knows the corresponding phase differences $z_{a+1}$, $\dots$, $z_{a+k-1}$.
Then in the second step, state $|\Phi\rangle$ of Eqn.(\ref{eq:defPhi}) sent to Bob
evolves inside his interferometer as
\begin{eqnarray}
|\Phi\rangle&\to&\frac12\sum_{i=-\infty}^\infty\left[(-1)^{y_i}A_i+(-1)^{y_{i-1}}A_{i-1}\right]d_{0i}^\dagger|0\rangle\label{eq:interferometer}\\
&&\ +\frac12\sum_{i=-\infty}^\infty\left[-(-1)^{y_i}A_i+(-1)^{y_{i-1}}A_{i-1}\right]d_{1i}^\dagger|0\rangle,\nonumber
\end{eqnarray}
where creation operators $d_{0i}^\dagger$ and $d_{1i}^\dagger$ denotes the single photons
incident to the detectors outputting $z_i=0$ and $1$ respectively.

An error in $z_i$ occurs when the spectrum $d_{z_i+1,i}^\dagger|0\rangle$
contained in Eqn.(\ref{eq:interferometer}) is measured by Bob.
Hence Bob detects error with probability
\[
E(k)
=\frac14\sum_{i=-\infty}^\infty\left|(-1)^{x_i+y_i}A_i-(-1)^{x_{i-1}+y_{i-1}}A_{i-1}\right|^2.
\]
By summing this over for Eve's random choices of $y_i$ for $i<a$ and $a+k\le i$,
we have
\begin{eqnarray}
E(k)&=&\frac14\sum_{i=a+1}^{a+k-1}|A_i-A_{i-1}|^2\label{eq:error_rate}\\
&&+\frac14\left(\sum_{i=-\infty}^a+\sum_{i=a+k}^\infty\right)\left[|A_i|^2+|A_{i-1}|^2\right].\nonumber
\end{eqnarray}

On the other hand,
Eve ends up knowing Alice and Bob's raw key bit when Bob detects either one
of phase differences $z_{a+1}$,
$\dots$,$z_{a+k-1}$, which occurs with probability
\begin{equation}
D(k)=\frac12\sum_{i=a+1}^{a+k-1}\left[|A_i|^2+|A_{i-1}|^2\right].
\label{eq:detection_rate}
\end{equation}
Note here that we are justified in including error events in $D(k)$
because the corresponding sifted key bit will always match between Eve and Bob
due to  key reconciliations to be performed later,
as long as the detected error rate is smaller than the threshold QBER.

\subsection{Effective Error Rate}
Hence in order for Eve to carry out the attack it is sufficient that there exist parameters
$1\le k$ and $0\le r\le1$ such that
\begin{eqnarray}
rp_{\rm click}&\le& p_{\rm seq}(k,\bar{n}),\label{eq:condition1}\\
rE(k)&\le& e_{\rm exp},\label{eq:condition2}
\end{eqnarray}
where $e_{\rm exp}$ is the QBER in the absence of Bob.
Inequality (\ref{eq:condition1}) means that Eve has an enough number of sequential events,
and Inequality (\ref{eq:condition2}) is to guarantee that the error rate is small enough so that
the presence of Eve will not be noticed by Bob.
And as a result of the attack,
Eve steals $rD(k)Np_{\rm click}$ bits out of an sifted key of length $Np_{\rm click}$ bits on average
while the error rate measured by Bob is $rE(k)$.

Since key reconciliation and the generalized privacy amplification are used in the
secret key generation of the DPSQKD \cite{BBCM95,Waks06},
the key generation rate $R$ is bounded from above as
\begin{eqnarray*}
R&\le&p_{\rm click}\left(I(A;B)-I(A;E)\right)\\
&=&p_{\rm click}\left(1-H_2(e_{\rm exp})-rD(k)\right),
\end{eqnarray*}
where $I(A;B)$ and $I(A;E)$ are the mutual informations between Alice and Bob,
and between Alice and Eve, respectively.
Clearly, $I(A;B)$ equals $1-H_2(e_{\rm exp})$
with $H_2(x)$ being the binary entropy function $H_2(x):=-x\log_2 x-(1-x)\log_2(1-x)$,
and $I(A;E)$ is given by $rD(k)$ in our sequential attack.

Thus if Eve is not interested in obtaining as many sifted key bits as possible,
but rather she is only willing to invalidate the key distribution between Alice and Bob,
the best strategy for her is to minimize the error rate $rE(k)$ by suppressing the ratio $r$.
That is, when Inequalities (\ref{eq:condition1}) and (\ref{eq:condition2}) are satisfied
for $r=1$ and a certain value of $k$,
the error rate $rE(k)$ measured by Bob can be further reduced by taking $r$ such that
\begin{equation}
1-H_2\left(rE(k)\right)-rD(k)=0.
\label{eq:def_effective}
\end{equation}
In what follows,
we will call such minimum value of $rE(k)$ the effective error rate and denote it by $E_{\rm eff}(k)$.

\section{Gaussian Intensity Modulation}\label{sec:gaussian}
In the original form of sequential attacks as introduced by Waks et al.\cite{Waks06},
the amplitude $A_i$s were
\[
A_i=\left\{
\begin{array}{cl}
1/\sqrt{k}&{\rm for}\ a\le i\le a+k-1,\\
0&{\rm otherwise},
\end{array}
\right.
\]
for which the error and detection rates were
\begin{equation}
E(k)=1/2k,\ {\rm and}\ D(k)=(k-1)/k.
\label{eq:prob_conventional}
\end{equation}
In this section we demonstrate how these can be improved by selecting an appropriate wave form for $A_i$.

One can generally seek for the best attack strategy, i.e.,
the best pattern of $A_i$ with the smallest $E(k)$ possible for each value of $k$
while maintaining the enough ratio $D(k)$ of bits accessible to Eve,
as long as $k$s are sufficiently small.
Indeed it is not difficult at all to find out the best $A_i$, e.g., by doing numerical simulations,
but in this paper, as the first trial,
we will concentrate on the following (discretized) gaussian pattern:
\begin{equation}
A_i=C\exp\left[-\frac{(i-c)^2}{4\sigma^2}\right].
\label{eq:def_gaussian}
\end{equation}
The constant $C$ appearing on the right hand side normalizes $|\Phi\rangle$ as $\sum_n|A_n|^2=1$,
$c$ denotes the time offset,
and $\sigma$ is the standard deviation to be adjusted later.
For the present we will set $c$ at the center of Eve's detected signals,
i.e., when Eve has detected phases $x_a,\cdots,x_{a+k-1}$,
we set $c=a+(k-1)/2$.

\subsection{First-Order Approximation of $E(k)$ and $D(k)$}
In order to have a rough idea as to how effective the gaussian intensity modulation in Eqn.(\ref{eq:def_gaussian}) is,
we first investigate it in the continuous limit of $k,\sigma\to\infty$,
where $E(k)$ and $D(k)$ given in Eqn.(\ref{eq:error_rate}) and (\ref{eq:detection_rate}) can be
explicitly calculated.
In this limit $C\to (\sqrt{2\pi}\sigma)^{-1/2}$ and we have
\begin{eqnarray*}
E(k)&=&\frac{|C|^2}4\sum_{i=-\infty}^{\infty}\left|A_i-A_{i-1}\right|^2\\
&&\ +\frac{|C|^2}2\left(\sum_{i=-\infty}^{a}+\sum_{i=a+k}^\infty\right)A_iA_{i-1}\\
&=&\frac{|C|^2}2\sum_{i=-\infty}^{\infty}\left(e^{\frac{-(i-c)^2}{2\sigma^2}}
-e^{-\frac{(i-c-1/2)^2}{2\sigma^2}-\frac1{8\sigma^2}}\right)\\
&&\ -|C|^2 e^{-\frac1{8\sigma^2}}\sum_{i=a+k}^\infty e^{-\frac{(i-c-1/2)^2}{2\sigma^2}}\\
&\simeq&\frac12\left(1-e^{-1/8\sigma^2}\right)
+\frac{e^{-1/8\sigma^2}}{\sqrt{2\pi}}\int_{k/2\sigma}^{\infty}dx\ e^{-x^2/2}\\
&\simeq&\frac1{16\sigma^2}+1-{\rm erf}\left(\frac{k}{2\sqrt{2}\sigma}\right)
\end{eqnarray*}
with ${\rm erf}(\cdot)$ being the error function
\[
{\rm erf}(x):=\frac2{\sqrt{\pi}}\int_0^xdt\ e^{-t^2}.
\]
Thus by choosing $k=4\sigma$ for example,
we have $E(k)\simeq 1/k^2+0.0228$, which is far smaller than Eqn.(\ref{eq:prob_conventional}).
Similarly, the detection rate can be calculated as
\[
D(k)\simeq\frac1{\sqrt{2\pi}}\int_{-k/2\sigma}^{k/2\sigma}dx\ e^{-x^2/2}={\rm erf}\left(\frac{k}{2\sqrt{2}\sigma}\right).
\]

\subsection{Corrections Due to Discretization}
In reality, we must take into account the corrections due to dicretization.
Adjusting values of $\sigma$,
we numerically calculated for each value of $k$ the smallest of error rate $E(k)$ as shown in Table \ref{table1}.
These values agree with the above first-order approximation within 10\% for $k\ge8$.
\begin{table}[ht]
\begin{center}
\begin{tabular}{|ccc|c|c|c|c|}\hline
\ \ &$k$&\ \ &  $\sigma$&\ \  $E(k)$\ \ &\ \ $D(k)$\ \ &\ \ $E_{\rm eff}(k)$\ \ \\ \hline
&4&&\ 0.871&0.105&0.881&0.0614\\
&5&&1.03&0.0748&0.930&0.0495\\
&6&&1.18&0.0562&0.954&0.0405\\
&7&&1.34&0.0438&0.968&0.0335\\
&8&&1.49&0.0353&0.977&0.0282\\
&9&&1.63&0.0290&0.982&0.0239\\
&10&&1.78&0.0243&0.987&0.0206\\
\hline
\end{tabular}
\caption{The smallest error rate $E(k)$ caused by $|\Phi\rangle$ with
the gaussian intensity modulation given in Eqn.(\ref{eq:def_gaussian}).
These minimums are obtained by adjusting the standard deviation $\sigma$ for each value of $k$.
Here $k$ denotes the length of Eve's sequential detection,
$D(k)$ the corresponding detection rate,
and $E_{\rm eff}(k)$ the effective error rate measured by Bob.}
\label{table1}
\end{center}
\end{table}
\section{Comparisons with Experiments}
In this section,
we discuss how our attack limits communication distances of the DPSQKD experiments.
The main purpose here is to show that in many of practical experimental setups,
our attack works more effectively than the individual attack introduce by Waks et al.\cite{Waks06},
hence reevaluating the security of several experiments reported up to now.

We will first show that the experiment with a communication distance of 100km
reported by Diamanti et al.\cite{Diamanti06} is in fact insecure.
Then we will also show that for a set of parameter values which are commonly used in 
today's QKD experiments and security analysis, e.g. in Ref.\cite{Diamanti06,Waks06},
the DPSQKD cannot be secure for distances larger than 95km no matter how one adjusts
the average photon number $\bar{n}$.

\subsection{100km Experiment by Diamanti et al.}
Diamanti et al.\cite{Diamanti06} performed a DPSQKD experiment with communication distance of 100km
using the  following set of parameters:
\begin{itemize}
\item[-] Average photon number $\bar{n}=0.2$,
\item[-] Optical fiber of 100km = loss of 20dB,
\item[-] Bob's Mach-Zehnder interferometer = loss of 2dB,
\item[-] Efficiency of Bob's detector\ =\ $4\times10^{-3}$,
\item[-] Bob's dark count probability $d=3.5\times10^{-8}$,
\end{itemize}
from which Bob's signal and dark count rate $p_{\rm click}$ of Eqn.(\ref{eq:detection_prob})
can be calculated as
\[p_{\rm click}=5.12\times10^{-6}\]
and the QBER they measured was
\[e_{\rm exp}=3.4\%.\]
With these values, however,
Eve can mount our sequential attack with $k=9$ and invalidate the secret key distribution.
The conditions (\ref{eq:condition1}) and (\ref{eq:condition2}) for $k=9$ are satisfied since
$p_{\rm click}<p_{\rm seq}(k=9,\bar{n}=0.2)=3.08\times10^{-5}$ as can be found from Eqn.(\ref{eq:sequential_det}),
and $E_{\rm eff}(9)=2.39\%<e_{\rm exp}$ from Table \ref{table1}.

\subsection{Limitations on the Communication Distance for a Practical Setup}
Next we compare our result with a more general set of experimental parameters given in Refs.\cite{Diamanti06} and \cite{Waks06}.

The signal and the dark count detection probability $p_{\rm click}$ is as given in (\ref{eq:detection_prob}),
where the transmission $T$ typically takes the form
\[
T=\eta_{\rm int}\eta_{\rm det}10^{-\alpha_{\rm fiber}L/10}.
\]
Here $\alpha_{\rm fiber}=0.2{\rm dB/km}$ is the fiber attenuation, $L$ the fiber length,
$\eta_{\rm int}=2{\rm dB}$ the loss in Bob's Mach-Zehnder interferometer\cite{Diamanti06},
and $\eta_{\rm det}=0.1$ the quantum efficiency of Bob's detector\cite{Waks06}.

On the other hand, the error rate in the absence of Eve is given by
\begin{equation}
e_{\rm exp}=\frac{\mu p_{\rm click}+d/2}{p_{\rm click}},
\label{eq:error_exp}
\end{equation}
where $\mu$ is the baseline error rate of the system due to imperfections in the state preparation,
channel induced noise, and imperfect detection apparatus \cite{Waks06}.
The typical values for $\mu$ and the dark count $d$ are $\mu=0.01$ and $d=10^{-5}$\cite{Waks06}.

Now let the communication distance $L=95{\rm km}$, or $T=10^{-3.1}$ (loss of 31dB).
Then it can be shown that key distribution is shown to be impossible for any value of $\bar{n}$ as follows;
For $\bar{n}\le0.14$ and $0.36\le\bar{n}$,
the gain formula $R_{\rm ind}$ based on individual attack\cite{Waks06} yields negative values:
\begin{eqnarray*}
R_{\rm ind}&=&-p_{\rm click}\left[(1-2\bar{n})\log_2P_{C_0}+H_2(e_{\rm exp})\right],\\
P_{C_0}&\le&1-e_{\rm exp}^2-\frac{(1-6e_{\rm exp})^2}2.
\end{eqnarray*}
On the other hand, for $\bar{n}\le0.14$ and $0.36\le\bar{n}$,
Inequalities (\ref{eq:condition1}) and (\ref{eq:condition2}) are satisfied for
values of $k$ given in Table \ref{table2}.
For example, 
Eve can mount the attack for $k=9$ when $0.30\le\bar{n}\le0.36$
because in this parameter region,
$p_{\rm seq}(9,\bar{n})-p_{\rm click}$ and $e_{\rm exp}(\bar{n},T)$ are both monotonically increasing in $\bar{n}$,
and it holds that for $\bar{n}=0.30$,
$p_{\rm seq}(9,\bar{n}=0.30)>p_{\rm click}(T,\bar{n}=0.30)$ and
$e_{\rm exp}(T,\bar{n}=0.30)=0.0302>E_{\rm eff}(9)=0.239$.
The effectiveness in other regions can be shown similarly.

\ 

\begin{table}[t]
\begin{center}
\begin{tabular}{|ccc|c|ccc|}\hline
&$\bar{n}$& &\ Attack Type\ \ &\ \ &$k$&\ \ \\ \hline
0&$\sim$&$0.14$\ \ &individual& &--&\\
$\ 0.14$&$\sim$&$0.19$\ \ &sequential& &6&\\
$\ 0.19$&$\sim$&$0.25$\ \ &sequential& &7&\\
$\ 0.25$&$\sim$&$0.30$\ \ &sequential& &8&\\
$\ 0.30$&$\sim$&$0.36$\ \ &sequential& &9&\\
$\ 0.36$&$\sim$&$\infty$&individual& &--&\\
\hline
\end{tabular}
\caption{The list of suitable attack type for a 95km DPSQKD experiment
depending on parameter regions of the average photon number $\bar{n}$.
The corresponding length $k$ of a sequential detection event are also shown in the right column.}
\label{table2}
\end{center}
\end{table}

\section{Conclusion}
In this paper,
we presented an improved version of sequential attacks with intensity modulation,
which works more effectively than the individual attack.
Using this attack, we have shown that the recent experiment with communication distance of 100km
reported by Diamanti et al.\cite{Diamanti06} is in fact insecure.
We also showed that in a practical experimental setup which is commonly used today,
the communication distance achievable by the DPSQKD protocol is less than 95km.

There are several straightforward ways to improve our result.
First, although we restricted ourselves in this paper to the discretized gaussian distribution
of Eqn.(\ref{eq:def_gaussian}),
numerically optimizing $A_i$ can yield a better form of state $|\Phi\rangle$ with
smaller error rate $E(k)$ while maintaining detection rate $D(k)$,
as mentioned at the beginning of Sec.\ref{sec:gaussian}.
Moreover, by letting $|\Phi\rangle$ be a superposition of states with different photon numbers,
we can increase the detection counts caused by each $|\Phi\rangle$ to more than one.
For instance, instead of the single-photon state $|\Phi\rangle$ given in Eqn.(\ref{eq:defPhi}),
it is possible that a coherent state
\[
|\Phi_{\rm coh}\rangle=\exp\left(\sum_i(-1)^{y_i}A_ia_i^\dagger+{\rm h.c.}\right)|0\rangle
\]
leads to a more effective attack.
Here h.c. stands for hermite conjugate.
Note that the norm $\sum_i|A_i|^2$ of $A_i$ corresponds to the average photon number contained in this state
and need not be normalized.

\ 

\noindent{\large\bf Acknowledgment}

This work was supported by the project ``Research and Development on Quantum Cryptography"
of the NICT as part of MPHPT of Japan's program ``R\&D on Quantum Communication Technology."

\end{document}